\documentclass{article}

\usepackage{arxiv}

\usepackage[utf8]{inputenc} 
\usepackage[T1]{fontenc}    
\usepackage[hidelinks]{hyperref}       
\usepackage{url}            
\usepackage{booktabs}       
\usepackage{amsfonts}       
\usepackage{nicefrac}       
\usepackage{microtype}      
\usepackage{lipsum}		
\usepackage{graphicx}
\usepackage{natbib}
\usepackage{doi}

\usepackage{url}
\usepackage{booktabs}
\usepackage{adjustbox,makecell,color,amsmath,multirow}

\title{Algebraic Connectivity Reveals Modulated High-Order Functional Networks \\in Alzheimer’s Disease}

\author{Giorgio Dolci$^{1,2}$, Silvia Saglia$^2$, Lorenza Brusini$^2$, Vince D. Calhoun$^3$, Ilaria Boscolo Galazzo$^2$, \\\textbf{Gloria Menegaz$^2$, for the Alzheimer's Disease Neuroimaging Initiative$^*$}\\\\
$^1$Department of Computer Science, University of Verona, Verona, Italy\\
$^2$Department of Engineering for Innovation Medicine, University of Verona, Verona, Italy\\
$^3$Tri-Institutional Center for Translational Research in Neuroimaging and Data Science, \\Georgia State University, Georgia Institute of Technology, Emory University, Atlanta, GA, USA.\\\\
$^*$Data used in this article were obtained from the Alzheimer’s Disease Neuroimaging Initiative (ADNI) database.\\The investigators within the ADNI contributed to the design and implementation of ADNI and/or provided data\\ but did not participate in the analysis/writing of this report.}

\date{}

\renewcommand{\shorttitle}{Algebraic Connectivity Enhances Hyperedge Specificity}

\begin{document}
\maketitle

\begin{abstract}
Functional MRI is a neuroimaging technique that analyzes the functional activity of the brain by measuring blood-oxygen-level-dependent signals throughout the brain. The derived functional features can be used for investigating brain alterations in neurological and psychiatric disorders. In this work, we employed a hypergraph to model high-order functional relations across brain regions, introducing algebraic connectivity ($a(\mathcal{G})$) for estimating the hyperedge weights.
The hypergraph structure was derived from healthy controls to build a common topology across individuals. The considered cohort for subsequent analyses included subjects covering the Alzheimer’s disease (AD) continuum, encompassing both mild cognitive impairment and AD patients. Statistical analysis and three classification tasks: HC vs AD, MCI vs AD, and HC vs MCI, were performed to assess differences across the three groups and the potential of the hyperedge weights as functional features.
Furthermore, a mediation analysis was performed to evaluate the reliability of the $a(\mathcal{G})$ values, representing functional information as the mediator between tau-PET levels, a key biomarker of AD, and cognitive scores. 
The proposed approach identified a larger number of hyperedges statistically different across groups compared to state-of-the-art methods. 
The $a(\mathcal{G})$ hyperedge weights also demonstrated a higher discriminative power in all three binary classifications.
Finally, two hyperedges belonging to salience/ventral attention and somatomotor networks showed a partial mediation effect between the tau biomarker and cognitive decline. These results suggested that $a(\mathcal{G})$ can be an effective approach for extracting the hyperedge weights, including important functional information that resides in the brain areas forming the hyperedges.
\end{abstract}


\section{Introduction}
magnetic resonance imaging (fMRI) is a widely used neuroimaging technique for studying changes in blood-oxygenation-level-dependent (BOLD) signals in the brain. The BOLD signals reflect the activation or deactivation of the brain areas over time, thus allowing the study of synchronous and asynchronous regions' temporal dynamics \cite{johnson2012brain}. Functional connectivity (FC) measures can be derived starting from fMRI signals, typically computed as the Pearson correlation between the time series data of each pair of brain regions and represented as symmetric connectivity matrices. FC has been extensively employed in research for investigating brain functions and identifying functional alterations possibly associated with neurodegenerative and neuropsychiatric disorders. Typically, these studies employ FC matrices in conjunction with common statistical techniques or machine/deep learning approaches to extract meaningful information \cite{alorf2022multi,geng2018multivariate,plitt2015functional}. 
From FC matrices, graph structures are extracted and used to investigate the network's patterns.
The graph is a data structure that, in the neuroscience field, can represent the brain, where each node of the graph represents a brain region, and the link between nodes is expressed as a measure of the association, typically the Pearson correlation \cite{gu2025fc,cui2022braingb}. Even if graph-based structures, corresponding metrics, and learning methods can better represent brain functions and modulations, they still represent an approximation of a complex system due to the pairwise relationships expressed by these data structures.

This bottleneck can be naturally overcome by switching to hypergraphs.
A hypergraph is a data structure consisting of hyperedges gathering multiple nodes. This enables the capture of high-order relations across different brain regions, providing a more flexible representation with a higher number of degrees of freedom. The hyperedges reflect functionally synchronized regions over time, representing distinct local functional subnetworks that go beyond the common functional networks \cite{gu2017functional}.
By grouping different regions under the same hyperedge, complex relations across brain areas can be captured, relying on ad-hoc measures for hyperedge shaping. 

Some works have already applied functional hypergraphs in neuroscience studies, e.g., for disease classification and brain functional analysis \cite{teng2024constructing,wang2024multiview,liu2023deep,gao2020hypergraph,xiao2019multi}. 

However, many issues remain open, leaving room for further research. 
The two main challenges are (i) building the hypergraph backbone structure, and (ii) assigning the weights to the hyperedges.
Regarding the first, various algorithms and techniques have been employed for detecting regions' interplay and creating hyperedges composed of highly related regions starting from the fMRI time series. Previous works relied on the k-nearest neighbor algorithm to define the hyperedges and corresponding brain regions \cite{teng2024constructing,zuo2021multimodal,zhou2006learning}. Other studies adopted clustering algorithms like k-means to learn the hypergraph structure \cite{ji2022fc,zhou2006learning}. The least absolute shrinkage and selection operator (LASSO) and Elastic Net, with their related extensions, are two other widely used methods that regress each time series with respect to all the others, hence extracting the functional relations across regions \cite{song2023brain,yang2023constructing,li2020hypernetwork,xiao2019multi,guo2018resting}. 
The second challenge regards weight computation. The hyperedge weights are crucial components of the data structure, as meaningful information can be extracted from these values, and subsequent analysis can be conducted based on them. 
A common approach for assigning weights to the hyperedges is the Gaussian similarity kernel. A distance measure is computed between pairs of node features or time series, and then the sum or mean inside the hyperedge is calculated \cite{niu2023applications,peng2023multiview,gao2020hypergraph,zhang2017joint}.
Another common approach to compute the hyperedge weights relies on the average or sum of the correlations across the regional time series of the same hyperedge \cite{bai2025hypergraph,song2023brain,shao2023analysis}. Other approaches derive the weights using a learning procedure, such as optimization procedures or deep learning algorithms \cite{wang2024multiview,ji2022fc,cai2023discovering,xiao2019multi}.

In this work, we propose a novel method to compute hyperedge weights, relying on the algebraic connectivity value computed over the brain regions that comprise the hyperedges as the metric to define the hyperedge weights. 
One of the reasons for choosing algebraic connectivity to weight the hyperedges is that this metric reflects the degree of connectivity in a graph \cite{de2007old}. Specifically, higher algebraic connectivity of a hyperedge reflects both higher synchronization and stronger network integration of its regions. On the contrary, lower values indicate poor synchronization and weaker network integration. From a biological perspective, a decrease in algebraic connectivity value can be associated with a loss of network robustness (e.g., the network becomes disconnected if a few nodes or links are removed) and reduced brain activity within the brain’s networks, deviating from the optimal configuration for dynamic processing \cite{daianu2014algebraic,de2012disruption}.
Hence, by comparing the same hyperedges across groups can reveal modulations in functional connectivity between subjects. Moreover, algebraic connectivity allows a fast and effective computation through an optimization algorithm. Indeed, it does not require a large amount of resources, representing a more convenient method compared to machine/deep learning models.
To evaluate this method, we analyzed fMRI data from participants along the Alzheimer’s disease (AD) spectrum, including healthy controls (HC), individuals with mild cognitive impairment (MCI), and patients with AD. We performed a group-level statistical and mediation analysis on the hyperedge weights to assess differences in FC across groups and to assess the mediation effect of the functional data between tau-PET levels, a key biomarker of AD etiopathogenesis \cite{jack2016t}, and cognitive decline. Additionally, three binary classification tasks were performed covering the full spectrum of AD to highlight the potential of the extracted hyperedge weights as functional features.

In summary, the aims of this work are threefold: (i) to derive hyperedge weights employing the algebraic connectivity measure in individuals across the AD continuum (HC, MCI, AD), based on a shared hypergraph structure; (ii) to assess its effectiveness in differentiating diagnostic groups in relation to high-order FC patterns through statistical analyses and machine learning binary classification tasks; and (iii) to examine whether hyperedge weights derived from algebraic connectivity mediate the relationship between tau burden and cognitive decline.

\section{Materials and Methods}
\subsection{Dataset}
Neuroimaging rs-fMRI data used in the preparation of this article were obtained from the Alzheimer's Disease Neuroimaging Initiative (ADNI) database (\url{https://adni.loni.usc.edu/}). The ADNI was launched in 2003 as a public-private partnership, led by Principal Investigator Michael W. Weiner, MD. The primary goal of ADNI was to test whether serial MRI, PET, other biological markers, and clinical and neuropsychological assessment can be combined to measure the progression of MCI and early AD.
\begin{table}[!htp]
    \centering
    \caption{Demographic information of the considered dataset.}
    \begin{adjustbox}{width=0.6\linewidth}
    \begin{tabular}{l l l l l}
        \toprule
         & \makecell[l]{HC-hypergraph} & \makecell[l]{HC-analysis} & \makecell[l]{MCI} & \makecell[l]{AD} \\
        \midrule
        Subjects & 217 & 93 & 199 & 78 \\
        Age & $73.9 \pm 8.7$ & $70.9 \pm 6.5$ & $73.8 \pm 11.0$ & $76.7 \pm 8.1$ \\
        Sex (M/F) & 102/115 & 33/60 & 109/90 & 47/31 \\
        Education & $16.9 \pm 2.3$ & $16.8 \pm 2.3$ & $16.3 \pm 2.6$ & $15.5 \pm 2.5$ \\
        ADAS-13 & $8.2 \pm 4.5$ & $7.8 \pm 4.9$ & $15.0 \pm 6.6$ & $30.1 \pm 8.6$ \\
        \bottomrule
    \end{tabular}
    \end{adjustbox}
    \label{tab:demos_info}
\end{table}

A total of 587 subjects were selected from the Phase 3 of ADNI (ADNI-3) and stratified along the AD continuum based on their clinical status, resulting in 310 HC, 199 MCI, and 78 AD. A subset of 217 HC subjects over the 310 HC was employed only for computing the hypergraph structure, named HC-hypergraph, and not for the subsequent analyses. In contrast, the remaining 93 HC (HC-analysis) individuals were used as a separate set only for the post-hoc analysis along with MCI and AD groups. Hence, the last three groups were not used to construct the hypergraph backbone.
Table \ref{tab:demos_info} shows the demographic information of the considered groups.

Rs-fMRI was acquired with the following acquisition protocol: TR/TE = 3000/30 ms, FA = 90°, FOV = 220 $\times$ 220 $\times$ 163 mm$^3$, 3.4-mm isotropic voxel size. 200 rs-fMRI volumes were acquired in almost all subjects, with minimal variations in a small subset (e.g., 197 or 195 volumes). More details on data acquisition can be found in \cite{weiner2017alzheimer}.

Standard preprocessing was applied to rs-fMRI, including removal of the first five volumes, head motion and slice-timing correction, nuisance regression (six motion parameters plus their derivatives, mean white matter (WM)/CSF signals and a linear trend component), band-pass filtering [0.01-0.08] Hz, linear plus non-linear spatial normalization to the 2-mm MNI space and smoothing with a 6-mm FWHM kernel \cite{pini2025functional}.
Then, the mean regional time series of the Schaefer 100 atlas \cite{yeo2011organization} were extracted, hence obtaining 100 time series with 192 time points for each subject. Finally, each time series was z-scored, resulting in a zero mean and unitary standard deviation.

\subsection{Hypergraph}
Brain network analysis can be performed using graph-based data structures, such as simple graphs and hypergraphs.
A simple graph $\mathcal{G}$ is a data structure that relies on pairwise relations and can be represented as $\mathcal{G}=(\mathcal{V},\mathcal{E},\mathbf{w})$, where $\mathcal{V}$ represents the set of vertices (nodes), $\mathcal{E}$ is the set of edges (connections between nodes), and $\mathbf{w}$ is the vector weights where each value is associated to each edge. Simple graphs have been widely used and investigated in brain functional analyses thanks to their ability in relating pairs of regions, observing synchronous and asynchronous activities, and studying brain modulations and alterations in neurodegenerative and neuropsychiatric disorders \cite{stam2024hub,stam2014modern}. Despite their potential and widespread use, they have an important limitation when dealing with complex systems, like the human brain. In detail, simple graphs can model only pairwise relations, thus limiting the understanding of brain functionalities and behavior. To solve this limitation, the hypergraph, which is a complex data structure, can be suitable since each hyperedge that defines the hypergraph can connect more than two nodes. A hypergraph $\mathcal{HG}$ is defined as $\mathcal{HG}=(\mathcal{V},\mathcal{E},\mathbf{W})$, where the vertex set $\mathcal{V}=\{v_1,v_2,...,v_N\}$ is composed of N vertices, the hyperedge set $\mathcal{E}=\{e_1,e_2,...,e_M\}$ is composed of M hyperedges, and $\mathbf{W}$ is the diagonal matrix of the hyperedge weights, where the $m$-th element in the diagonal corresponds to weight of the $m$-th hyperedge, $\mathbf{W}=diag(\{w(e_1),w(e_2),...w(e_M)\}$.
The hypergraph can be represented by an incidence matrix $\mathbf{H} \in \mathbb{R}^{N \times M}$, where
\begin{equation}\label{eq:incidence_matrix}
    \mathrm{H}_{n,m} = 
    \left\{ 
    \begin{array}{l}
        1 \quad\mathrm{if}\ v_n \in e_m \\
        0 \quad\mathrm{otherwise}.
    \end{array} 
\right.
\end{equation}
Additionally, the incidence matrix $\mathbf{H}$ can also be non-binary, expressed as a weighted incidence matrix composed of real values (not only zeros and ones) in order to assign a specific importance to each node in the different hyperedges. 
For each node, its degree represents the importance of the node in the $\mathcal{HG}$  across the hyperedges. Similarly, the hyperedge degree represents the number of nodes belonging to the hyperedge. This information in the hypergraph can be expressed in terms of diagonal matrices and, building on this, the node and hyperedge degree diagonal matrices ($\mathbf{D}_v \in \mathbb{R}^{N \times N}$ and $\mathbf{D}_e \in \mathbb{R}^{M \times M}$, respectively) can be defined as follows:
\begin{align}
    d(v_n) &= \sum_{e_m \in \mathcal{E}}w(e_m)\mathrm{H}_{n,m} &\mathrm{for}\ 1\le n \le N\\
    \delta(e_m) &= \sum_{v_n \in \mathcal{V}}\mathrm{H}_{n,m} &\mathrm{for}\ 1\le m \le M
\end{align}
where $d(v_n)$ and $\delta(e_m)$ represent the diagonal values for each node and hyperedge of $\mathbf{D}_v$ and $\mathbf{D}_e$, respectively. Finally, a hypergraph similarity matrix $\mathbf{S} \in \mathbb{R}^{N \times N}$ is defined as
\begin{equation}
    \mathbf{S} = \mathbf{HW}\mathbf{D}_{e}^{-1}\mathbf{H}^{T} \label{eq:simm_matrix}
\end{equation}
where the elements of the matrix $\mathbf{S}$ represent high-order relations across the nodes of the hypergraph $\mathcal{HG}$ weighted by the hyperedge weights. 

\subsubsection{Hypergraph Structure Extraction}
The $\mathcal{HG}$ structure in the form of the incidence matrix $\mathbf{H}$ was derived using LASSO regression, a well-established method for hypergraph structure computation \cite{jie2016hyper,li2019multimodal,xiao2019multi}. 
LASSO regression was computed in order to identify relationships between regional time series, considering the $n$-th time series (region $n$) as dependent variable and the other $N$-1 time series as independent variables. Therefore, for each time series $n$ in $1\le n \le N$, the optimization problem is defined as
\begin{align}
    & \underset{\boldsymbol{\alpha}_n}{\mathrm{min}} \frac{1}{2}||\mathbf{x}_n - \mathbf{X}_{n}\boldsymbol{\alpha}_n||_2^2\ +\ \lambda||\boldsymbol{\alpha}_n||_1 \\
    &\mathrm{subject\ to}\quad \boldsymbol{\alpha}_n \succeq 0
\end{align}
where the constraint prevents the negative coefficients, $\mathbf{x}_n$ is the $n$-th time series used as the response variable, $\mathbf{X}_{n} \in \mathbb{R}^{P \times N-1}$ is the time series matrix of each subject, excluding the $n$-th time series (replaced by a zero vector), $\boldsymbol{\alpha}_n$ corresponds to the coefficients relating the response region with the other brain areas, and $\lambda$ is the sparsity hyperparameter used to extract a sparse representation of the $\boldsymbol{\alpha}_n$ vector. The $\lambda$ was optimized through 5-fold cross validation, testing different values in the range of [0.01, 0.30], with a step of 0.01. Specifically, for every $\lambda$, for every subject, and region used as centroid, the 5-fold cross validation was performed on the timepoints of the fMRI time series. This allowed to find the LASSO model that best reconstructed the centroid using the other regions as features. At the end of this procedure, $\lambda$ = 0.05 was selected based on the overall best performance, where the average mean squared error across all subjects and regions was $0.115 \pm 0.02$ for training and $0.257 \pm 0.05$ for validation.

The LASSO regression coefficients were extracted for all the centroid regions ($n$-th region), resulting in a coefficient matrix $\mathbf{A}_{s} \in \mathbb{R}^{N \times N}$ for each subject $s$.
Then, the distribution of the values in each $\mathbf{A}_{s}$ was computed and subsequently thresholded, considering the fifth percentile to remove the smallest and non-relevant coefficients. As a further step, all $\mathbf{A}_{s}$ were binarized.
After that, only the hyperedges with at least three brain regions, including the centroid region, were considered for the subsequent analysis, thus obtaining an incidence matrix $\mathbf{H}_{s} \in \mathbb{R}^{N \times M}$ for each subject. The threshold of a minimum of three regions was selected to consider all the possible high-order interactions while avoiding the pairwise connections captured by conventional graph analyses. All the extracted hyperedges were considered undirected since the regions that formed them were directly linked with the centroid region, and the same centroid region was part of the hyperedge.

Finally, to extract a common $\mathcal{HG}$ structure for all the individuals, the majority voting values across all the incidence matrices $\mathbf{H}_{s}$ were computed. In detail, for each cell, the value of 1 or 0 was selected if at least 50\% of the HC-hypergraph individuals had a value of 1 or 0, respectively. The result consisted of a matrix $\mathbf{H}$ that represented the $\mathcal{HG}$ structure in the form of incidence matrix, as described in Eq. (\ref{eq:incidence_matrix}). 

\begin{figure}[!htp]
    \centering
    \includegraphics[width=0.45\linewidth]{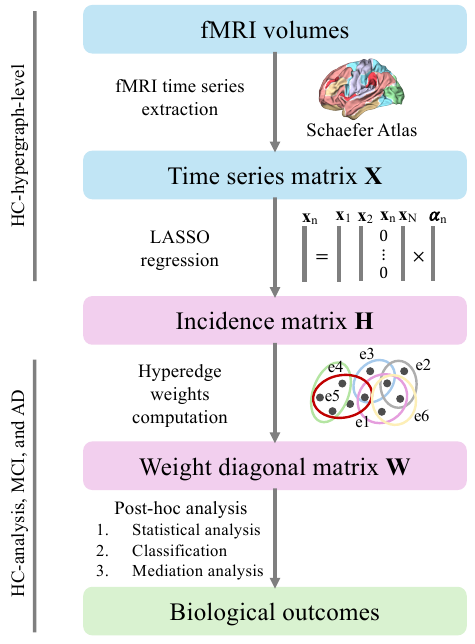}
    \caption{Pipeline of this work. Initially, the mean time series are extracted from the fMRI volumes for each region of the Schaefer atlas. Then, the LASSO regression is computed between the intra-subject time series to extract the coefficients' matrix, subsequently binarized in the incidence matrix $\mathbf{H}$. After that, the hyperedge weights (matrix $\mathbf{W}$) are computed. Finally, a post-hoc analysis is performed.}
    \label{fig:pipeline}
\end{figure}

\begin{figure}
    \centering
    \includegraphics[width=0.5\linewidth]{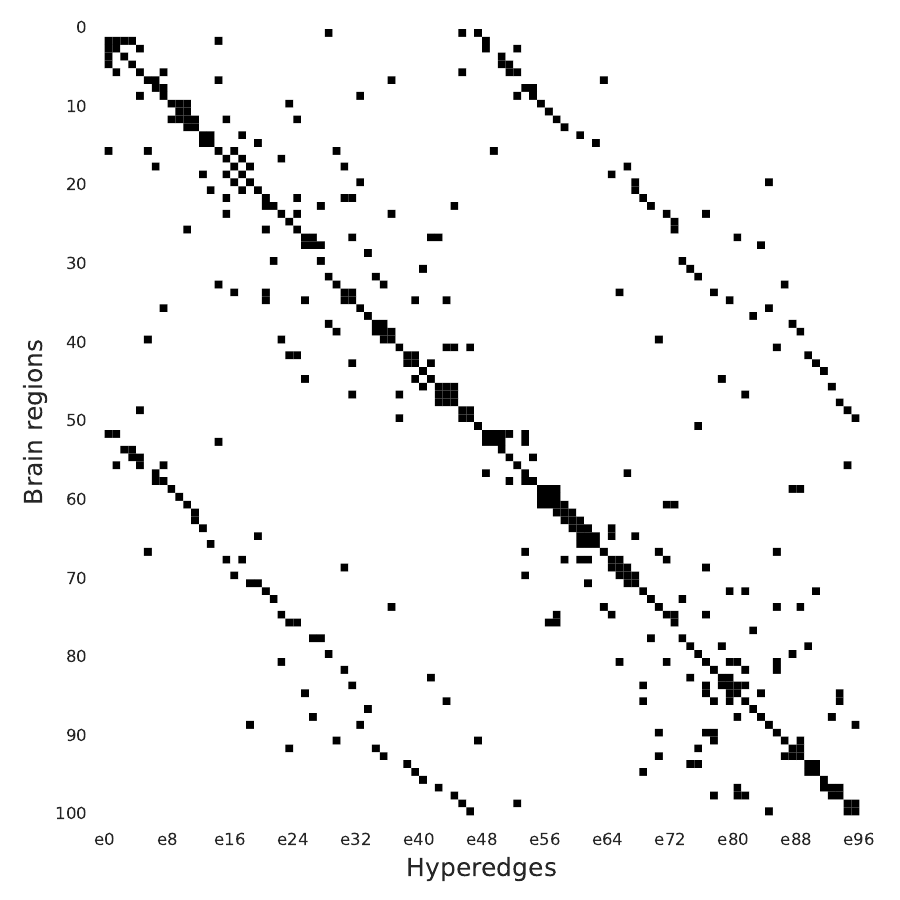}
    \caption{Incidence matrix H of the hypergraph. H is composed of 96 hyperedges, with a maximum of 7 regions per hyperedge, a minimum of 3, and an average of 4.66 regions.}
    \label{fig:incidence_mat}
\end{figure}

Figure \ref{fig:pipeline} shows the pipeline used to extract the hypergraph structure, while Fig. \ref{fig:incidence_mat} shows the incidence matrix $\mathbf{H}$ of the hypergraph. The resulting hypergraph $\mathcal{HG}$ is composed of 96 hyperedges with a maximum of seven regions per hyperedge, a minimum of three, and an average of 4.66 regions. The cardinality of each hyperedge was calculated using Eq. (3), which extracts the hyperedge degree. The extracted $\mathcal{HG}$ structure remained stable even if shorter time series were employed, by truncating the final portion of each fMRI signal. This stability was assessed using the Jaccard index and Pearson's correlation between the $\mathcal{HG}$ structure extracted from the full-length time series and the shorter signals.

\subsubsection{Hyperedge Modeling}
Different methods have been proposed for computing the hyperedge weights ($\mathbf{W}$ matrix). The $\mathbf{W}$ matrix is a fundamental aspect when working with hypergraphs, since via this information different features can be extracted, such as the node degree matrix $\mathbf{D}_v$ and the hypergraph similarity matrix $\mathbf{S}$.
To this end, in this work, we proposed a novel method of computing the weight of each hyperedge, relying on the algebraic connectivity ($a(\mathcal{G})$, Fiedler value) \cite{fiedler1973algebraic}. The importance of $a(\mathcal{G})$ resides in the fact that it is a measure of how well connected a graph is \cite{de2007old}. The $a(\mathcal{G})$ is computed on the subgraph composed of the regions belonging to a specific hyperedge. In detail, the absolute value of the FC matrix defined by the regional time series belonging to the hyperedge was computed, thus generating a subgraph of the brain regions of the hyperedge. The $a(\mathcal{G})$ is defined as the second-smallest eigenvalue extracted from the Laplacian matrix of the subgraph. Due to the limited number of regions involved in each hyperedge (at least three), the trace minimization with LU factorization approach was used to decompose the Laplacian matrix, and the tolerance value for computing the eigenvalues was set to $1e-10$ in order to have numerical stability when calculating the $a(\mathcal{G})$.

\subsection{Post-hoc Analysis}
\subsubsection{Baseline Approaches}
The $a(\mathcal{G})$ approach was compared via a post-hoc analysis with three common approaches to compute the hyperedge weights, and with the mean correlation value of the Schaefer 100 functional networks (FNs), where each FN was considered as a hyperedge. In detail:
\paragraph*{Gaussian similarity kernel} The Gaussian similarity kernel, defined as $K[i,j] = \exp{\frac{-||\mathbf{x}_i - \mathbf{x}_j||_2^2}{2\sigma^2}}$, was computed across the time series of the regions belonging to the hyperedge, where $\mathbf{x}_i$ and $\mathbf{x}_j$ are the $i$-th and $j$-th time series, respectively, and $\sigma$ was set using the heuristic median technique.

\paragraph*{Mean Pearson correlation} The mean correlation was computed across all pairwise correlations calculated between pairs of time series belonging to the hyperedge.

\paragraph*{L2 norm of LASSO coefficients} The L2 norm of the LASSO coefficients in the hyperedge in the form of $\mathbf{w}_m = ||\boldsymbol{\alpha}_m||_2^2$.

\paragraph*{Mean FN correlation} The mean correlation values calculated from the seven FNs of the Schaefer 100 atlas were employed as the hyperedge weights, where the hyperedges were expressed in terms of FNs regions.

\subsubsection{Statistical Analysis} \label{sub:stats_analysis}
A statistical analysis was performed on the HC-analysis, MCI, and AD individuals, relying on the weights calculated for all the hyperedges. In detail, for each hyperedge, the non-parametric Kruskal-Wallis test was performed to identify the hyperedges that were statistically different across the three groups in terms of hyperedge weights. The non-parametric test was selected due to the non-normal distribution of the data assessed through the Shapiro-Wilk test. 

For each significant hyperedge, the Cliff's $\delta$ effect size was computed in absolute value between group pairs and then averaged to obtain a global effect size across the three groups.

Finally, a post-hoc analysis was performed to highlight the differences between the group pairs employing the Mann-Whitney test on the hyperedges that resulted in significant differences from the previous statistical test. The false discovery rate (FDR) strategy was used to correct the p-values for both analyses.

In order to check for any bias introduced using two different groups of HC for the hypergraph construction and for the post-hoc analysis, the same statistical analysis was also performed between the HC-hypergraph and HC-analysis individuals.

\subsubsection{Binary Classifications}
Three binary classification tasks have been modeled not to outperform the state-of-the-art in AD classification, but to assess the potential of the extracted functional features from the hypergraph to discriminate between the different clinical groups. To this end, the statistically different hyperedge weights derived from each method have been used, separately, as input features for a random forest model to classify, respectively, (i) HC vs AD, (ii) MCI vs AD, and (iii) HC vs MCI. For each task, an initial train-test split (80/20) was performed, using the training set to tune the random forest hyperparameters and train the model using 5-fold cross validation, while retaining the test set as a hold-out set for the test phase. The same train/test sets were used for all methods employed to compute the hyperedge weights, but each random forest was optimized independently. The models' performance was reported in terms of accuracy, precision, recall, and F1 score for the hold-out test set. Finally, the internal features importance of the random forest models was extracted to understand which hyperedge contributed most to the classification.

\subsubsection{Mediation Analysis}
Entorhinal cortex, a small region in the medial temporal lobes, is among the earliest sites affected by AD pathology and is considered a key origin point for tau accumulation \cite{igarashi2023entorhinal}. From this condition, pathological tau spreads to other brain regions via neural connections, contributing to disease progression and cognitive decline. The spread of tau pathology in AD is believed to follow a prion-like mechanism, advancing transneuronally across synaptically and functionally connected regions \cite{wang2024characterization}. At the same time, tau deposition is known to disrupt FC, suggesting a reciprocal relationship between network integrity and tau progression \cite{roemer2025amyloid, franzmeier2022tau}. 
Motivated by this interplay, we conducted a mediation analysis, with and without covariates (age, sex, years of education, and APOE4), to explore whether high-order functional features captured by hypergraph-derived hyperedge weights of all methods might mediate the link between entorhinal cortex tau burden, as measured by SUVR from tau-PET, and cognitive performance. The same analysis was performed on the weights extracted from the Schaefer 100 FNs, where each FN was considered as a hyperedge. To note, this analysis was conducted on a subset of the original cohort of ADNI-3 and considering all diagnostic groups jointly. More specifically, five MCI and three AD patients did not have the entorhinal cortex tau SUVR values and/or cognitive scores available, and for this reason, they were excluded from the analysis.

\begin{figure}[!hbp]
    \centering
    \includegraphics[width=0.45\linewidth]{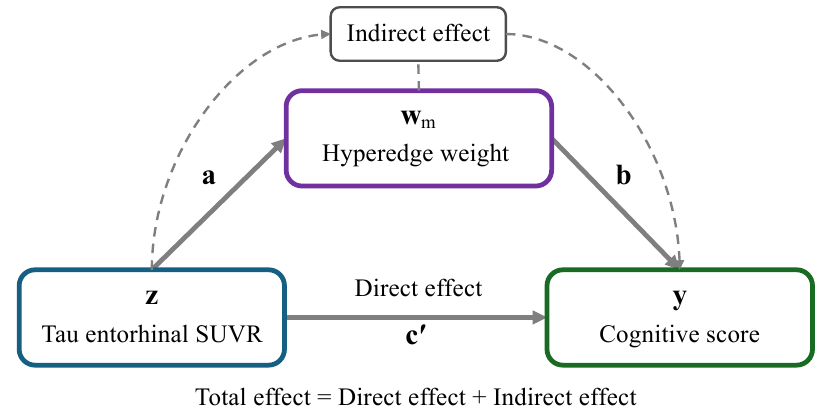}
    \caption{Schematic representation of the mediation analysis of this work.}
    \label{fig:mediation_analysis}
\end{figure}

The following linear regressions modeled the relations between variables:
\begin{align}
    \mathbf{y} &= \mathbf{i}_1 + \mathbf{cz} + \mathbf{e}_1 \\
    \mathbf{w}_m &= \mathbf{i}_2 + \mathbf{az} + \mathbf{e}_2 \\
    \mathbf{y} &= \mathbf{i}_3 + \mathbf{c}'\mathbf{z} + \mathbf{bw}_m + \mathbf{e}_3
\end{align}
where $\mathbf{y}$ represents the cognitive composite scores (memory, executive, language, visual-spatial, and ADAS-13), $\mathbf{z}$ the tau SUVR in entorhinal cortex, $\mathbf{w}_m$ is the mediator variable, corresponding to the weight of the $m$-th hyperedge under analysis (and the FNs weights). $\mathbf{i}_{1,2,3}$, $\mathbf{e}_{1,2,3}$, and $\mathbf{c,a,b,c}'$ are the intercepts, residual errors, and coefficients of the regression models, respectively. The $\mathbf{c}'$ and $\mathbf{ab}$ define the direct and indirect effects of the mediation analysis, respectively, where the direct effect represents the effect of tau on the cognitive score without mediator (functional information), while the indirect one is the effect of tau SUVR on cognitive scores through the functional information (mediator). Figure \ref{fig:mediation_analysis} shows the graphical representation of the mediation analysis.

\begin{figure*}
    \centering
    \includegraphics[width=0.9\linewidth]{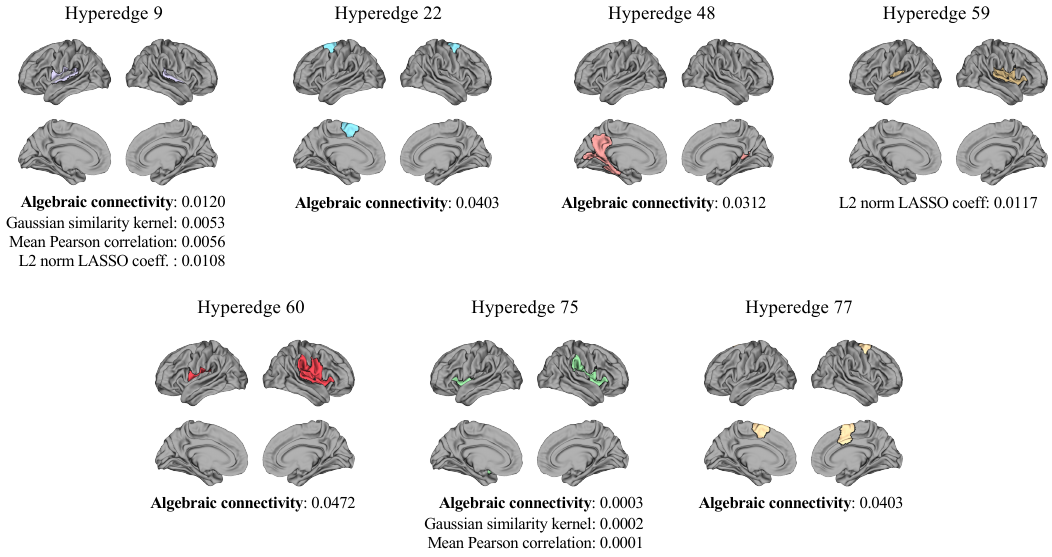}
    \caption{Hyperedges found statistically relevant in the proposed analysis with their FDR-corrected p-values for the four different approaches (FDR level of 0.05). The highlighted regions were mainly part of salience/ventral attention, somatomotor, dorsal attention, default-mode, and visual FNs.}
    \label{fig:hyperedges_regions}
\end{figure*}

\begin{figure*}
    \centering
    \includegraphics[width=\linewidth]{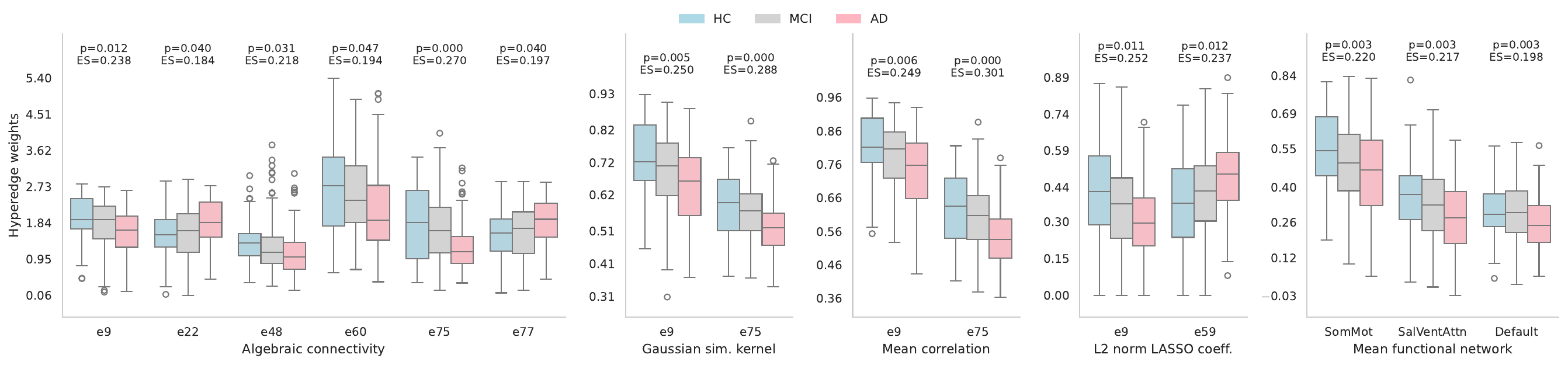}
    \caption{Boxplots representing the distribution of the weights of the three groups for the statistically significant hyperedges and FNs. In the figure, the hyperedges are denoted by the letter "e" followed by the corresponding hyperedge number used in this study. For each hyperedge, the Kruskal-Wallis FDR-corrected p-value ($\mathrm{p}<0.05$) and corresponding effect size (ES) are annotated.}
    \label{fig:boxplots_sign_hyperedges}
\end{figure*}

All the hyperedges' weights extracted using the $a(\mathcal{G})$ and the baseline approaches that resulted in statistically significant differences across the three groups were employed for the mediation analysis. Additionally, a bootstrap procedure with 1000 repetitions was employed to assess the significance of the indirect effect of the mediation analysis.

\subsubsection{External Validation}
We tested the robustness of the $a(\mathcal{G})$ considering an atlas with higher resolution. We extracted the fMRI time series of the Schaefer 400 atlas from the fMRI volumes for all the subjects, resulting in 400 regional time series with 192 time points each. The hypergraph backbone was extracted using the same HC-hypergraph individuals employed for defining the hypergraph structure of the Schaefer 100 atlas. Finally, considering the same cohort of HC-analysis, MCI, and AD subjects from the previous analyses, we conducted the statistical analyses described in \ref{sub:stats_analysis} for all hyperedge weight computation methods.

We additionally tested the potential of the $a(\mathcal{G})$ on a different cohort of subjects from Phase 2 of ADNI (ADNI-2), who did not overlap with the selected cohort of ADNI-3. A total of 96 individuals were collected, including 21 HC (Age: $75.8\pm 8.1$; Sex: M=9/F=12; MMSE: $28.9\pm1.7$), 42 MCI (Age: $73.5\pm 8.6$; Sex: M=22/F=20; MMSE: $27.5\pm 2.1$), and 33 AD (Age: $74.6\pm 6.7$; Sex: M=15/F=18; MMSE: $21.3\pm 4.5$). We considered the same hypergraph backbone computed on the HC-hypergraph subjects of ADNI-3 based on the Schaefer 100 atlas, but statistical analyses across the three groups and for post-hoc pairwise tests as described in \ref{sub:stats_analysis} were performed on the new cohort.

\section{Results}
\subsection{Group Differences in Extracted Weights}
The $\mathcal{HG}$ structure was extracted using the HC-hypergraph group for the subsequent analyses. The stability of the resulting $\mathcal{HG}$ structure was evaluated with respect to the length of the time series. Stable results were obtained for fMRI data comprising more than 135 time points. At this lower bound, comparison between $\mathcal{HG}$ structures extracted from signals with 135 and 192 time points yielded a Jaccard index of 0.855 and a correlation of 0.921. Consistently, the number of hyperedges was comparable, with 90 and 96 hyperedges for the 135 and 192 time points signals, respectively.

Based on these results, only the $\mathcal{HG}$ structure computed from the full-length time series was used in the subsequent analysis.
A non-parametric statistical analysis was performed on the hyperedge weights extracted from all the approaches to assess differences across HC, MCI, and AD subjects. 
Figure \ref{fig:hyperedges_regions} shows the brain regions forming hyperedges that exhibited statistically significant differences across groups. The corresponding FDR-corrected p-values are also reported. Table \ref{tab:hyperedges_info} summarizes the number of regions and their associated FNs for each significant hyperedge. Figure \ref{fig:boxplots_sign_hyperedges} illustrates the distribution of these hyperedge weights, along with their corresponding FDR-corrected p-values and effect sizes, stratified by group and weighting method. Results showed that hyperedges 9, 48, 60, and 75, which include regions from the somatomotor, visual, and default-mode FNs, exhibited reduced $a(\mathcal{G})$ in AD compared to both HC and MCI. In contrast, hyperedges 22 and 77, composed of regions within the dorsal and salience/ventral attention FNs, showed increased $a(\mathcal{G})$ in AD patients. Similarly, the other approaches highlighted the same patterns with higher values in HC than MCI/AD for hyperedges 9 and 75. Further, the global Cliff's $\delta$ effect size revealed a small separation effect across the three groups, with the higher values for hyperedge 75.

The weights computed with the $a(\mathcal{G})$ highlighted a larger number of significant hyperedges compared to the other three approaches, identifying six hyperedges with significant differences. The mean Pearson correlation, the L2 norm of the LASSO coefficients, and the Gaussian similarity kernel identified only two significant hyperedges. All four methods highlighted differences across groups for hyperedge 9, which comprises regions within the somatomotor FN. Hyperedge 75, composed of four regions from the salience/ventral attention FN and one region from the somatomotor FN, was identified as significant by three of the methods, including the $a(\mathcal{G})$. When examining the FNs hyperedge weights, three out of seven FNs were found to be statistically different: the somatomotor, salience/ventral attention, and default-mode FNs.

\begin{table*}[!htp]
    \centering
    \caption{List of regions involved in the statistically different hyperedges. For each region, the region name of the Schaefer 100 atlas, the MNI centroid coordinates, the hyperedge containing the region, and the main overlap in percentage with the anatomical regions of the Harvard-Oxford atlas are reported. The FNs are included in the region name (Vis=Visual; SomMot=Somatomotor; DorsAttn=Dorsal attention; SalVentAttn=Salience/Ventral attention; Default=Default-mode), while LH and RH mean left and right hemisphere. Post. and ant. means posterior and anterior division.}
    \begin{adjustbox}{width=\textwidth}
    \begin{tabular}{l l l l}
        \toprule
        \makecell[l]{Region name} & \makecell[l]{MNI Centroid Coordinates} & \makecell[l]{Hyperedges} & \makecell[l]{Main overlaps (Harvard-Oxford atlas)} \\
        \midrule
        LH Vis 1 & (-26, -34, -18) & 48 & Parahippocampal Gyrus, post. (62\%); Temporal Fusiform Cortex, post. (22\%) \\
        LH Vis 6 & (-12, -66, 6) & 48 & Intracalcarine Cortex (53\%); Lingual Gyrus (11\%) \\
        LH SomMot 1 & (-54, -22, 8) & 9 & Heschl's Gyrus (with H1 and H2) (35\%); Planum Temporale (32\%) \\
        LH SomMot 2 & (-36, -22, 16) & 59 & Insular Cortex (31\%); Central Opercular Cortex (24\%) \\
        LH SomMot 3 & (-54, -12, 14) & 9, 60 & Central Opercular Cortex (68\%) \\
        LH DorsAttn FEF 1 & (-26, -4, 60) & 22 & Superior Frontal Gyrus (33\%); Middle Frontal Gyrus (15\%) \\
        LH SalVentAttn FrOperIns 1 & (-42, -2, -6) & 75 & Insular Cortex (64\%) \\
        LH SalVentAttn FrOperIns 2 & (-38, 12, 6) & 75 & Frontal Operculum Cortex (59\%); Central Operculum Cortex (21\%) \\
        LH SalVentAttn Med 3 & (-6, 4, 62) & 22, 77 & Supplementary Motor Cortex (49\%) \\
        LH Default pCunPCC 1 & (-12, -56, 12) & 48 & Precuneous Cortex (44\%) \\
        LH Default pCunPCC 2 & (-6, -52, 34) & 48 & Precuneous Cortex (50\%); Cingulate Gyrus (42\%) \\
        RH SomMot 1 & (52, -16, 6) & 9, 59, 60 & Heschl's Gyrus (with H1 and H2) (43\%); Planum Temporale (12\%) \\
        RH SomMot 2 & (40, -16, 16) & 59, 60 & Central Opercular Cortex (41\%); Insular Cortex (24\%) \\
        RH SomMot 3 & (56, -4, 12) & 59, 60, 75 & Central Opercular Cortex (48\%) \\
        RH SomMot 4 & (58, -6, 30) & 60 & Postcentral Gyrus (39\%); Precentral Gyrus (29\%) \\
        RH DorsAttn FEF 1 & (28, -2, 60) & 22, 77 & Superior Frontal Gyrus (38\%); Middle Frontal Gyrus (21\%) \\
        RH SalVentAttn TempOccPar 2 & (60, -26, 28) & 60, 75 & Supramarginal Gyrus, ant. (49\%); Parietal Operculum Cortex (19\%) \\
        RH SalVentAttn FrOperIns 1 & (40, 8, 2) & 59, 60, 75 & Insular Cortex (50\%); Central Opercular Cortex (13\%) \\
        RH SalVentAttn Med 2 & (8, 6, 52) & 77 & Supplementary Motor Cortex (62\%) \\
        RH Default pCunPCC 1 & (12, -54, 14) & 48 & Precuneous Cortex (56\%) \\
        \bottomrule
    \end{tabular}
    \end{adjustbox}
    \label{tab:hyperedges_info}
\end{table*}

To further assess pairwise group differences, post-hoc comparisons were performed using the Mann-Whitney test. Table \ref{tab:posthoc_stat_analysis} summarizes the results of this analysis, highlighting the number of hyperedges that exhibited significant differences between each pair of groups (HC vs MCI, HC vs AD, and MCI vs AD).
\begin{table}[!htp]
    \centering
    \caption{Post-hoc analysis showing the number of significant hyperedges between group pairs over the total number of hyperedges found in each method. The approaches could discriminate the HC-AD and MCI-AD groups, but not HC-MCI.}
    \begin{adjustbox}{width=0.6\linewidth}
    \begin{tabular}{l c c c}
    \toprule
                                 & HC vs AD & HC vs MCI & MCI vs AD \\
    \cmidrule{2-4}
    Algebraic connectivity       & 6/6      & 0/6       & 6/6       \\
    Gaussian similarity kernel   & 2/2      & 0/2       & 2/2       \\
    Mean Pearson correlation     & 2/2      & 0/2       & 2/2       \\
    L2 norm LASSO coeff.         & 2/2      & 1/2       & 2/2       \\
    Functional networks          & 3/3      & 1/3       & 2/3       \\
    \bottomrule
    \end{tabular}
    \end{adjustbox}
    \label{tab:posthoc_stat_analysis}
\end{table}
The post-hoc analysis revealed a consistent ability across all considered approaches to discriminate between HC and AD, as well as MCI and AD. On the other hand, no hyperedge weights were able to differentiate the HC and MCI groups, except for hyperedge 9 identified by the LASSO-based method. Similarly, the somatomotor FNs of the Schaefer 100 atlas allowed differentiating HC and MCI groups.

Finally, the control analysis performed between the HC-hypergraph and HC-analysis found no differences between the two groups for all the methods, except for the Gaussian similarity kernel approach, which identified one hyperedge with an FDR-corrected $p < 0.05$.

\subsection{Classification Results}
\begin{table}[!htp]
    \centering
    \caption{Classification results for the three tasks: (i) HC vs AD (top), (ii) MCI vs AD (center), and (iii) HC vs MCI (bottom).}
    \begin{adjustbox}{width=0.5\linewidth}
    \begin{tabular}{l l l l l}
        \toprule
        \makecell[l]{HC vs AD} & \makecell[l]{ACC} & \makecell[l]{PRE} & \makecell[l]{REC} & \makecell[l]{F1} \\
        \midrule
        Algebraic connectivity     & \textbf{0.800} & \textbf{0.831} & \textbf{0.800} & \textbf{0.798} \\
        Gaussian similarity kernel & 0.657 & 0.656 & 0.657 & 0.654 \\
        Mean Pearson correlation           & 0.771 & 0.771 & 0.771 & 0.771 \\
        L2 norm LASSO coef.        & 0.657 & 0.663 & 0.657 & 0.658 \\
        \midrule\\

        \makecell[l]{MCI vs AD} & \makecell[l]{ACC} & \makecell[l]{PRE} & \makecell[l]{REC} & \makecell[l]{F1} \\
        \midrule
        Algebraic connectivity     & \textbf{0.750} & \textbf{0.750} & \textbf{0.750} & \textbf{0.691} \\
        Gaussian similarity kernel & 0.696 & 0.506 & 0.696 & 0.586 \\
        Mean Pearson correlation           & 0.679 & 0.503 & 0.679 & 0.578 \\
        L2 norm LASSO coef.        & 0.714 & 0.510 & 0.714 & 0.595 \\
        \midrule\\

        \makecell[l]{HC vs MCI} & \makecell[l]{ACC} & \makecell[l]{PRE} & \makecell[l]{REC} & \makecell[l]{F1} \\
        \midrule
        Algebraic connectivity     & \textbf{0.695} & \textbf{0.687} & \textbf{0.695} & 0.609 \\
        Gaussian similarity kernel & \textbf{0.695} & 0.666 & \textbf{0.695} & 0.644 \\
        Mean Pearson correlation           & 0.644 & 0.586 & 0.644 & 0.593 \\
        L2 norm LASSO coef.        & 0.678 & 0.645 & 0.678 & \textbf{0.644} \\
        \bottomrule
    \end{tabular}
    \end{adjustbox}
    \label{tab:classification_results}
\end{table}
Three classification tasks were addressed, relying on the hyperedge weights as input features for the random forest model for each hyperedge weight computation method. The hyperedge weights extracted using the $a(\mathcal{G})$ showed a more discriminative power in the three tasks, except for the HC vs MCI comparisons, where comparable performance was obtained by the Gaussian similarity kernel approach. In detail, in the first task (HC vs AD), $a(\mathcal{G})$ achieved an accuracy of 0.800, while the Gaussian similarity kernel, mean correlation, and the L2 norm of the LASSO coefficients obtained accuracies of 0.657, 0.771, and 0.657, respectively. For the second task (MCI vs AD), accuracies of 0.750, 0.696, 0.679, and 0.714 were achieved by $a(\mathcal{G})$, Gaussian similarity kernel, mean correlation, and L2 norm of LASSO coefficients, respectively, showing how the $a(\mathcal{G})$ outperformed the other methods. Finally, the model on the third task (HC vs MCI) yielded the same result of 0.695 using the $a(\mathcal{G})$ as well as the Gaussian similarity kernel. In contrast, 0.644 and 0.678 were obtained by mean correlation and L2 norm of the LASSO coefficients, respectively. Table \ref{tab:classification_results} shows the performance of the methods for the three tasks, reporting accuracy, precision, recall, and F1 score.

The importance of the features of the different models across the tasks revealed that hyperedge 75 was the most important for $a(\mathcal{G})$. Similarly, hyperedge 75 was relevant for the Gaussian similarity kernel and mean correlation methods in discriminating MCI from AD and HC from MCI, while hyperedge 9 was relevant for distinguishing the HC group from AD. Finally, hyperedge 59 was the most important in all tasks for the L2 norm of the LASSO coefficients.

\subsection{Mediation Analysis Results}
\begin{table*}[!htp]
    \centering
    \caption{Statistically significant results of the mediation analysis with the bootstrap method (significance set at $p < 0.05$) examining the relationship across tau entorhinal cortex SUVR, algebraic connectivity values, and five cognitive composite scores, showing both results without (top) and with covariates (age, sex, years of education,
    and APOE4) (bottom). Values represent the coefficients of the regression forming the total, direct, and indirect effects, with their corresponding confidence intervals [2.5\%,97.5\%] and p-value (p-val).}
    \begin{adjustbox}{width=\textwidth}
    \begin{tabular}{c r r r r r r r r r r r}
    \toprule
    Hyperedge & \multicolumn{11}{c}{Mediation analysis for Algebraic connectivity - $a(\mathcal{G})$ - without covariates} \\
    \cmidrule{2-12}
     & & \multicolumn{2}{c}{Memory} & \multicolumn{2}{c}{Executive} & \multicolumn{2}{c}{Language} & \multicolumn{2}{c}{Visual-Spatial} & \multicolumn{2}{c}{ADAS-13} \\
    \cmidrule{2-12}
     & & Coeff & p-val & Coeff & p-val & Coeff & p-val & Coeff & p-val & Coeff & p-val \\
    \multirow{5}{*}{60} & $\mathrm{w_m}\sim\mathrm{z}$ & -0.108 [-0.211,-0.005] & 0.0394 &  &  & -0.108 [-0.211,-0.005] & 0.0394 &  &  & -0.108 [-0.211,-0.005] & 0.0394 \\
    
     & $\mathrm{y}\sim\mathrm{w_m}$ & 0.178 [ 0.096, 0.260] & $<$0.001 & & & 0.148 [ 0.087, 0.210] & $<$0.001 & & & -2.228 [-3.255,-1.202] & $<$0.001\\
    
     & Total & -0.479 [-0.548,-0.412] & $<$0.001 & & & -0.214 [-0.274,-0.154] & $<$0.001 & & & 6.246 [ 5.417, 7.076] & $<$0.001\\
    
     & Direct & -0.466 [-0.533,-0.399] & $<$0.001 & & & -0.200 [-0.259,-0.142] & $<$0.001 & & & 6.076 [ 5.257, 6.896] & $<$0.001\\
    
     & Indirect & -0.014 [-0.034,-0.001] & 0.046 & & & -0.014 [-0.031, 0.000] & 0.046 & & & 0.170 [ 0.008, 0.409] & 0.048\\

    & & & & & & & & & & & \\
    
    \multirow{5}{*}{75} & $\mathrm{w_m}\sim\mathrm{z}$ & -0.192 [-0.294,-0.090] & $<$0.001 & -0.192 [-0.294,-0.090] & $<$0.001 & -0.192 [-0.294,-0.090] & $<$0.001 & -0.192 [-0.294,-0.090] & $<$0.001 & -0.192 [-0.294,-0.090] & $<$0.001\\
    
     & $\mathrm{y}\sim\mathrm{w_m}$ & 0.199 [ 0.117, 0.280] & $<$0.001 & 0.158 [ 0.083, 0.233] & $<$0.001 & 0.136 [ 0.074, 0.199] & $<$0.001 & 0.106 [ 0.041, 0.172] & 0.002 & -2.684 [-3.699,-1.699] & $<$0.001\\
    
     & Total & -0.479 [-0.548,-0.412] & $<$0.001 & -0.303 [-0.373,-0.233] & $<$0.001 & -0.214 [-0.274,-0.154] & $<$0.001 & -0.164 [-0.228,-0.099] & $<$0.001 & 6.246 [ 5.417, 7.076] & $<$0.001\\
    
     & Direct & -0.459 [-0.527,-0.390] & $<$0.001 & -0.283 [-0.354,-0.212] & $<$0.001 & -0.195 [-0.255,-0.135] & $<$0.001 & -0.149 [-0.214,-0.084] & $<$0.001 & 5.950 [ 5.119, 6.781] & $<$0.001\\
    
     & Indirect & -0.021 [-0.043,-0.008] & $<$0.001 & -0.019 [-0.042,-0.007] & 0.006 & -0.019 [-0.036,-0.007] & $<$0.001 & -0.015 [-0.031,-0.005] & 0.008 & 0.296 [ 0.130, 0.559] & $<$0.001\\

    \midrule

    Hyperedge & \multicolumn{11}{c}{Mediation analysis for Algebraic connectivity - $a(\mathcal{G})$ - with covariates} \\
    \cmidrule{2-12}
     & & \multicolumn{2}{c}{Memory} & \multicolumn{2}{c}{Executive} & \multicolumn{2}{c}{Language} & \multicolumn{2}{c}{Visual-Spatial} & \multicolumn{2}{c}{ADAS-13} \\
    \cmidrule{2-12}
     & & Coeff & p-val & Coeff & p-val & Coeff & p-val & Coeff & p-val & Coeff & p-val \\
    \multirow{5}{*}{75} & $\mathrm{w_m}\sim\mathrm{z}$ & -0.192 [-0.306,-0.078] & $<$0.001 & -0.192 [-0.306,-0.078] & $<$0.001 & -0.192 [-0.306,-0.078] & $<$0.001 & -0.192 [-0.306,-0.078] & $<$0.001 & -0.192 [-0.306,-0.078] & $<$0.001\\
    
     & $\mathrm{y}\sim\mathrm{w_m}$ & 0.137 [ 0.061, 0.212] & $<$0.001 & 0.113 [ 0.040, 0.186] & 0.002 & 0.088 [ 0.028, 0.148] & 0.004 & 0.079 [ 0.013, 0.145] & 0.019 & -1.988 [-2.946,-1.031] & $<$0.001\\
    
     & Total & -0.457 [-0.527,-0.387] & $<$0.001 & -0.269 [-0.346,-0.193] & $<$0.001 & -0.213 [-0.275,-0.148] & $<$0.001 & -0.126 [-0.198,-0.053] & $<$0.001 & 5.934 [ 5.048, 6.819] & $<$0.001\\
    
     & Direct & -0.444 [-0.515,-0.373] & $<$0.001 & -0.255 [-0.332,-0.178] & $<$0.001 & -0.201 [-0.265,-0.137] & $<$0.001 & -0.114 [-0.187,-0.041] & 0.002 & 5.725 [ 4.833, 6.617] & $<$0.001\\
    
     & Indirect & -0.013 [-0.033,-0.001] & 0.046 & -0.014 [-0.037,-0.001] & 0.036 & -0.011 [-0.027,-0.001] & 0.004 & -0.012 [-0.031,-0.002] & 0.032 & 0.209 [ 0.059, 0.459] & 0.004\\
    
    \bottomrule
    \end{tabular}
    \end{adjustbox}
    \label{tab:mediation_analysis_results}
\end{table*}

Table \ref{tab:mediation_analysis_results} summarizes the results of the mediation analysis performed using $a(\mathcal{G})$ with and without covariates. Among the six hyperedges that showed significant group differences, only two revealed significant mediation effects without considering the covariates. Specifically, hyperedge 60 showed a partial mediation effect between tau and the memory, language, and ADAS-13 composite scores. Hyperedge 75 revealed a broader significant partial mediation effect, linking the tau accumulation with all the cognitive scores. Across the analysis, the regression coefficients showed a similar pattern in the two hyperedges. In detail, negative relationships were found between tau burden and memory, executive, language, and visual-spatial scores, where higher tau values were associated with cognitive decline, e.g., lower values in the composite cognitive scores. Conversely, a positive effect was observed for the ADAS-13 score, where higher values were associated with the disease. The same analysis was performed, controlling for the covariates. Only the weights of hyperedge 75 showed a significant partial mediation effect with all the scores. The same patterns observed in the regression coefficients for hyperedge 75 without covariates were also found after controlling for age, sex, years of education, and APOE4. In general, we also observed a reduction in the total effects of the mediation analysis when including the covariates, and consequently, the indirect effect was decreased with respect to the mediation analysis without covariates.

The Gaussian similarity kernel and the mean Pearson correlation revealed significant mediation analysis for the weights of hyperedge 75 and all the cognitive scores with/without covariates (except for the visual-spatial score where no mediation effect was found with mean Pearson correlation), while the L2 norm of the LASSO coefficients approach found only significant relations in hyperedge 9 and for the memory (with/without covariates) and ADAS-13 (without covariates) scores. No significant mediation effects were found when using the Schaefer 100 FNs-derived weights as mediators.

\subsection{External Validation Results}
The statistical analysis described in \ref{sub:stats_analysis} for the Schaefer 100 atlas was repeated using the high-resolution Schaefer 400 atlas to assess the robustness of $a(\mathcal{G})$. $a(\mathcal{G})$ identified five significant hyperedges after FDR correction, while the Gaussian similarity kernel, mean correlation, and L2-norm LASSO approaches identified six, zero, and one significant hyperedges, respectively. Although the Gaussian kernel found one additional hyperedge, $a(\mathcal{G})$ showed stronger discriminative power in post-hoc pairwise tests, yielding significant differences (FDR-corrected) for all hyperedges across all group comparisons. Conversely, the Gaussian similarity kernel identified five out of six hyperedges as significantly different for HC vs AD and MCI vs AD, and two out of six hyperedges for HC vs MCI. Furthermore, since the FDR correction was applied to a large number of hyperedges, we also investigated a less stringent significance threshold (p $<$ 0.10). Under this threshold, both $a(\mathcal{G})$ and the Gaussian similarity kernel identified the same number of hyperedges (10). The five additional hyperedges of $a(\mathcal{G})$ showed a trend towards significance. Specifically, their FDR-corrected p-values were close to 0.05, namely 0.054, 0.058, 0.058, 0.061, and 0.081. However, the effect size of the significant hyperedges was computed, and all the methods showed a small effect size across the three groups. Biologically, the five significant hyperedges identified by $a(\mathcal{G})$ were composed of regions mainly corresponding to the salience/ventral attention, default-mode, control, and somatomotor FNs. Instead, the five close-to-significance hyperedges were associated with the visual and dorsal attention FNs. Notably, the dorsal and salience/ventral attention, default-mode, somatomotor, and visual FNs were also relevant when considering the Schaefer 100 atlas.

Similarly, the same analyses described in \ref{sub:stats_analysis} were performed using the individuals of ADNI-2. None of the methods detected significant hyperedges after FDR correction. However, using raw p-values, $a(\mathcal{G})$ identified 10 hyperedges, while the Gaussian similarity kernel, the mean correlation, and the L2 norm of the LASSO coefficients detected a lower number (four with both the Gaussian similarity kernel and the mean correlation, and three with the L2 norm of the LASSO coefficients). Considering these hyperedges, in post-hoc pairwise comparisons, $a(\mathcal{G})$ demonstrated the ability to differentiate between group pairs (8/10 significantly different hyperedges for HC vs MCI, 7/10 for HC vs AD, and 2/10 for MCI vs AD). The Gaussian kernel detected mainly HC vs AD differences (0/4 for HC vs MCI, 4/4 for HC vs AD, 2/4 for MCI vs AD), mean correlation showed effects in HC vs MCI and HC vs AD (3/4 for HC vs MCI, 3/4 for HC vs AD, 0/4 for MCI vs AD), and LASSO highlighted changes across all pairs but most strongly in MCI vs AD (2/3 for HC vs MCI, 1/3 for HC vs AD, 3/3 for MCI vs AD). Additionally, two hyperedges found to be statistically different in the ADNI-2 cohort were also identified as significant in ADNI-3, considering $a(\mathcal{G})$. A small effect size was found for all methods across groups. In this case, the hyperedges were composed mainly of regions of the salience/ventral attention, dorsal attention, somatomotor, and default-mode FNs.

\section{Discussion}
The hypergraph is a data structure able to model high-order relations across brain regions, thereby better approximating and representing a more realistic nature of the brain. The key concepts of the hypergraph are related to the hyperedges and their weights, since they represent groups of highly related brain regions and the importance of the different hyperedges in the hypergraph, respectively.

Selecting the most appropriate method for computing the hyperedge weight is essential, since it carries the functional information that resides in the brain areas of that subgraph. Assigning constant values or setting all weights to ones can be too simplistic \cite{yang2021graphlshc,bai2021hypergraph} and should be avoided since no meaningful features can be extracted. Similarly, assigning the hyperedge degree as the weight \cite{pisarchik2025hypergraph} results in a partial loss of the regional functional data, keeping only the information related to the number of regions forming the hyperedges.

In this study, taking advantage of the brain functional data, more complex approaches have been considered and developed for computing the hyperedges' weights. Specifically, we explored approaches such as the Gaussian similarity kernel \cite{niu2023applications,peng2023multiview,gao2020hypergraph,zhang2017joint}, the L2 norm of LASSO coefficients \cite{balogh2022hypergraph,gao2020hypergraph,pan2021characterization}, the mean or sum of pairwise correlations values between time series of the regions forming each hyperedge \cite{bai2025hypergraph,song2023brain,shao2023analysis}, and the mean correlation values of the regions defined by the Schaefer 100 functional atlas. With these approaches, we identified the hyperedges 9 (somatomotor areas) and 75 (somatomotor and salience/ventral attention regions) and three FNs (somatomotor, salience, salience/ventral attention, and default-mode) that showed significant differences in AD vs MCI and AD vs HC, highlighting hypoconnectivity in the patients' groups with respect to the HC one. These findings were consistent across the baseline approaches, including the outcomes of the Schaefer 100 FNs, and with previous works that have observed hypoconnectivity in AD \cite{wiesman2021somatosensory,li2012attention}. On the other hand, a hyperconnectivity effect was featured in the hyperedge 59 (somatomotor and salience/ventral attention areas) found by the L2 norm-based method, where higher connectivity was found in AD.

Even if these approaches obtained relevant results concordant with previous works, they relied on averaging or summing pairwise metrics, thus not featuring the real nature of the hypergraph, except for the L2 norm applied to the LASSO coefficients that aggregated the regression coefficients. On these bases, the advantage of using the $a(\mathcal{G})$ derives from the fact that featuring the Laplacian matrix of the subgraph (hyperedge) and its decomposition to the second-smallest eigenvalue, we can extract a measure of how well the hyperedges are connected avoiding the use of mean or sum operations which brings approximation in the final results.
The proposed method identified a larger number of hyperedges (9, 22, 48, 60, 75, and 77, where 9 and 75 were in common with the other approaches) statistically different across groups. 
These hyperedges included regions of the somatomotor, dorsal and salience/ventral attention, default-mode, and visual FNs. 
Interestingly, some hyperedges (e.g., 48, 60, and 75) grouped regions belonging to different FNs, such as somatomotor and attention, and visual and default-mode, which are usually considered separately in standard atlases. This suggests that our approach with hypergraphs could capture more insights and functional organization in AD that conventional methods overlook. 
Most of the significant hyperedges showed reduced $a(\mathcal{G})$ in AD patients, consistent with the idea of a disease-related loss of functional integration \cite{pini2021breakdown}. 
This pattern appeared most prominently in hyperedges formed by regions belonging to the salience/ventral attention, somatomotor, visual, and default-mode FNs which have been previously linked to AD in earlier studies \cite{dolci2025interpretable,dolci2025multimodal,dai2019disrupted}. Specifically, prior works have observed functional disruptions in the salience/ventral attention and somatomotor FNs, indicating alterations in these systems during AD progression. Likewise, reduced functional integrity in the default-mode and visual FNs has been associated with AD, with greater declines in these FNs corresponding to more severe cognitive impairment \cite{dai2019disrupted,huang2021alzheimer,agosta2010sensorimotor,li2015toward,jalilianhasanpour2019functional}.
In contrast, hyperedges involving dorsal and salience/ventral attention FNs showed increased connectivity in the AD patients, reflecting possible compensatory effects. 
The compensatory effect pattern in these FNs is still debated in the literature. However, some studies found a similar trend to the one observed in our results, i.e., AD greater than MCI greater than HC in terms of functional power, thus confirming our proposed results \cite{franzmeier2017resting,jalilianhasanpour2019functional,balthazar2014neuropsychiatric}.

Further, $a(\mathcal{G})$ demonstrated a higher capability to address various classification tasks using the extracted features, i.e., the hyperedge weights. $a(\mathcal{G})$ outperformed the other approaches in the three classification tasks except for the most critical one (i.e., HC vs MCI), where similar performances to the Gaussian similarity kernel were obtained. Regarding the importance of the extracted features, hyperedge 75 was the most relevant hyperedge that guided the classifications, consistent with the larger effect size found across groups, in contrast to the other hyperedge weights.

Additionally, the relationship between tau pathology and cognitive impairment was partially mediated by the high-order functional information resided in hyperedges 60 and 75, suggesting possible modulations in the coordination of large-scale networks in AD\cite{iordan2024salience,sperling2019impact,bejanin2017tau,mielke2017association,jack2016t}. 
A similar pattern was previously observed employing the FC networks related to episodic memory and executive function as mediators between the tau levels and the memory and executive scores. They found that a significant mediation effect of FC was present between tau and memory function, but not in the executive \cite{ziontz2024behaviorally}. Their results partially agree with our findings, where, for the hyperedges 60 and 75, significant mediation effects were present for the memory score. On the contrary, only in the hyperedge 75 a mediation effect between tau and executive function due to FC was observed.

The potential of $a(\mathcal{G})$ was further supported using a higher-resolution atlas and an independent cohort. In general, $a(\mathcal{G})$ demonstrated stronger discriminative power in pairwise comparisons and across groups compared to the other methods, except for the Gaussian similarity kernel. Indeed, the Gaussian similarity kernel detected one additional statistically different hyperedge when an atlas at a higher resolution was employed. The biological composition of the hyperedges found by $a(\mathcal{G})$, considering both Schaefer 400 and an independent ADNI cohort, revealed FNs (salience/ventral and dorsal attention, default-mode, somatomotor, and visual) that were consistent with the results of Schaefer 100. Considering even higher resolution atlases or additional datasets with different acquisition parameters (e.g., scan length and TR) could be of interest for assessing further the stability of the framework and thus deserve further investigations, even if some attentions have to be taken. In particular, fine-grained atlases (e.g., 800-1000 parcels) could hamper a direct biological interpretation of the information that resides in the small parcels, and might introduce noisy features with possible unstable hyperedges.

Taken together, the obtained results highlighted the plausibility and potential of the $a(\mathcal{G})$ for deriving hyperedge weights that can capture meaningful functional information in the AD continuum and help understand the neurophysiological mechanisms underlying AD.

\subsection{Main Limitations}
One possible limitation of this work could be related to the LASSO method employed for computing the hypergraph structure. Even if this method has been widely used and the objective of this work was not to define new strategies for extracting the hypergraph structure, other approaches than LASSO could be considered for defining it. Another possible limitation could be related to the $a(\mathcal{G})$ computation. This approach requires the graph to be connected, otherwise it will result in a null score. Such a constraint limits this approach only to connected graphs. Even though no disconnected subgraphs were obtained in the current study, a feasible strategy is to discard those hyperedges from the analysis since no comparisons could be performed. Finally, even if $a(\mathcal{G})$ showed better results than the other methods, the Cliff's $\delta$ effect size found across the three groups was relatively small.

\section{Conclusion}
In this work, we proposed an approach for computing the hyperedge weights in the hypergraph structure relying on the $a(\mathcal{G})$, which is a measure of how well the brain regions composing the hyperedges are connected, even if they reside in different FNs. The results obtained with $a(\mathcal{G})$ showed a greater number of significantly different hyperedges and superior classification performance than common approaches. The combination of decreased connectivity in somatomotor, attention, and default-mode FNs, together with increased connectivity in attention systems, underscores the heterogeneous and dynamic nature of the disease. Our use of $a(\mathcal{G})$ and hypergraph models provides a new way to capture these changes, highlighting both functional disconnection and potential adaptations. Furthermore, the mediation effects involving tau burden reinforce the idea of FC as the mechanistic link between molecular pathology and clinical symptoms, offering a pathway through which disease processes affect cognition. 

\section*{Data Availability Statement}
The data used in this work were collected by ADNI (\url{https://adni.loni.usc.edu/}), and they are publicly available upon request on the ADNI website.

The code used for the analyses of this work can be found in this GitHub repository: \url{https://github.com/BraiNavLab/Algebraic_Connectivity_Hypergraph_Alzheimer.git}.

\section*{Acknowledgments}
Data collection and sharing for this project was funded by the Alzheimer's Disease Neuroimaging Initiative (ADNI) (National Institutes of Health Grant U01 AG024904) and DOD ADNI (Department of Defense award number W81XWH-12-2-0012). ADNI is funded by the National Institute on Aging, the National Institute of Biomedical Imaging and Bioengineering, and through generous contributions from the following: AbbVie, Alzheimer’s Association; Alzheimer’s Drug Discovery Foundation; Araclon Biotech; BioClinica, Inc.; Biogen; Bristol-Myers Squibb Company; CereSpir, Inc.; Cogstate; Eisai Inc.; Elan Pharmaceuticals, Inc.; Eli Lilly and Company; EuroImmun; F. Hoffmann-La Roche Ltd and its affiliated company Genentech, Inc.; Fujirebio; GE Healthcare; IXICO Ltd.; Janssen Alzheimer Immunotherapy Research \& Development, LLC.; Johnson \& Johnson Pharmaceutical Research \& Development LLC.; Lumosity; Lundbeck; Merck \& Co., Inc.; Meso Scale Diagnostics, LLC.; NeuroRx Research; Neurotrack Technologies; Novartis Pharmaceuticals Corporation; Pfizer Inc.; Piramal Imaging; Servier; Takeda Pharmaceutical Company; and Transition Therapeutics. The Canadian Institutes of Health Research is providing funds to support ADNI clinical sites in Canada. Private sector contributions are facilitated by the Foundation for the National Institutes of Health (\url{www.fnih.org}). The grantee organization is the Northern California Institute for Research and Education, and the study is coordinated by the Alzheimer’s Therapeutic Research Institute at the University of Southern California. ADNI data are disseminated by the Laboratory for Neuro Imaging at the University of Southern California.

\bibliographystyle{unsrt}
\bibliography{main}

\end{document}